\documentclass[runningheads]{llncs}
\usepackage[T1]{fontenc}
\usepackage{amsmath}

\usepackage{graphicx,verbatim}
\begin{document}
\title{Efficient Epistemic Uncertainty Estimation in Cerebrovascular Segmentation}
%
\author{Omini Rathore\and
Richard Dominik Paul\and
Abigail Morrison\and
Hanno Scharr\and 
Elisabeth Pfaehler}
%
%
\institute{Forschungszentrum Jülich, Jülich, Germany  }    
\maketitle              
\begin{abstract}
Brain vessel segmentation of MR scans is a critical step in the diagnosis of cerebrovascular diseases. Due to the fine vessel structure, manual vessel segmentation is time consuming. Therefore, automatic deep learning (DL) based segmentation techniques are intensively investigated. As conventional DL models yield a high complexity and lack an indication of decision reliability, they are often considered as not trustworthy. This work aims to increase trust in DL based models by incorporating epistemic uncertainty quantification into cerebrovascular segmentation models for the first time. By implementing an efficient ensemble model combining the advantages of Bayesian Approximation and Deep Ensembles, we aim to overcome the high computational costs of conventional probabilistic networks. Areas of high model uncertainty and erroneous predictions are aligned which demonstrates the effectiveness and reliability of the approach. We perform extensive experiments applying the ensemble model on out-of-distribution (OOD) data. We demonstrate that for OOD-images, the estimated uncertainty increases. Additionally, omitting highly uncertain areas improves the segmentation quality, both for in- and out-of-distribution data. The ensemble model explains its limitations in a reliable manner and can maintain trustworthiness also for OOD data and could be considered in clinical applications. 


\end{abstract}
\section{Introduction}
Stroke is a significant public health burden globally, with a particularly high incident rate in Asia ~\cite{ref_article1}. 
One cause of stroke is cerebrovascular disease affecting the brain's blood vessels. To diagnose cerebrovascular diseases, a three-dimensional vessel segmentation extracted from Magnetic Resonance Angiograms (MRA) or Computed Tomography Angiograms (CTA) images is essential and can be critical for treatment decisions and prognosis ~\cite{ref_proc1}.

Manual segmentation of brain vessels is labor-intensive, prone to inter-observer variability, and human error ~\cite{ref_article2}. To get fast and reproducible segmentation results, an automatic segmentation process is desirable. In recent years, a large number of research works have explored (semi-) automatic segmentation techniques  ~\cite{ref_proc1}. Hereby, Deep learning (DL)-based methods, such as convolutional neural networks (CNNs) have surpassed accuracy of traditional methods and are one possible candidate for clinical implementation.\\
Despite these advancements, barriers to adopt DL in healthcare persist. One reason for the missing clinical application is the black box nature of CNNs what hinders clinicians’ trust ~\cite{ref_article3}. Additionally, due to factors such as a limited amount of training data, a CNN may produce unreliable decisions, e.g. when a data point is presented that differs highly from the training data. To increase trust in DL models, an estimate on decision reliability is of uttermost importance. \\
Uncertainty quantification (UQ) can provide insights into the model’s confidence and has been examined alongside explainability in numerous studies to identify cases where a model lacks confidence ~\cite{ref_article4} ,~\cite{ref_corr}. UQ is especially interesting for medical images, where image quality can vary highly between data acquired at different scanners and/or with different settings. It can happen that a model trained on data from one center provides unreliable results when applied to images from another center. Indication of such unreliabilities is essential to allow clinicians to manually correct automated segmentations. Besides the advantages, a limitation of UQ methods is their high computational costs. Moreover, it is unclear how to handle voxels identified as uncertain model decisions. In this work, we aim to address these points by performing the following steps:
\begin{enumerate}
    \item Development of a 3D U-net model ~\cite{ref_proc2} for cerebrovascular segmentation of Time-of-Flight MRA images (TOF-MRA).
    \item Implementation of an Efficient Ensemble Model generation method ~\cite{ref_proc3} to estimate epistemic uncertainties. 
    \item Analysis of the correlation between uncertainty and specific patterns in cerebrovascular structures to provide insights into when and why the model’s predictions may be less reliable. 
    \item Demonstrating the use of uncertainty estimates in detecting out-of-distribution (OOD) data and in further improving the segmentation results. 
\end{enumerate}
\section{Materials and Methods}
\textbf{Efficient Ensemble Model}
Each network architecture can be used as backbone for the efficient ensemble model used in this work \cite{ref_proc3}. The efficient ensemble model consists of M sub-models differing in model initialization of the respective backbone architecture. From these sub-models, the ensemble members are dynamically constructed by stochastically selecting sub-model layers.
\begin{figure}
    \centering    \includegraphics[width=0.8\linewidth]{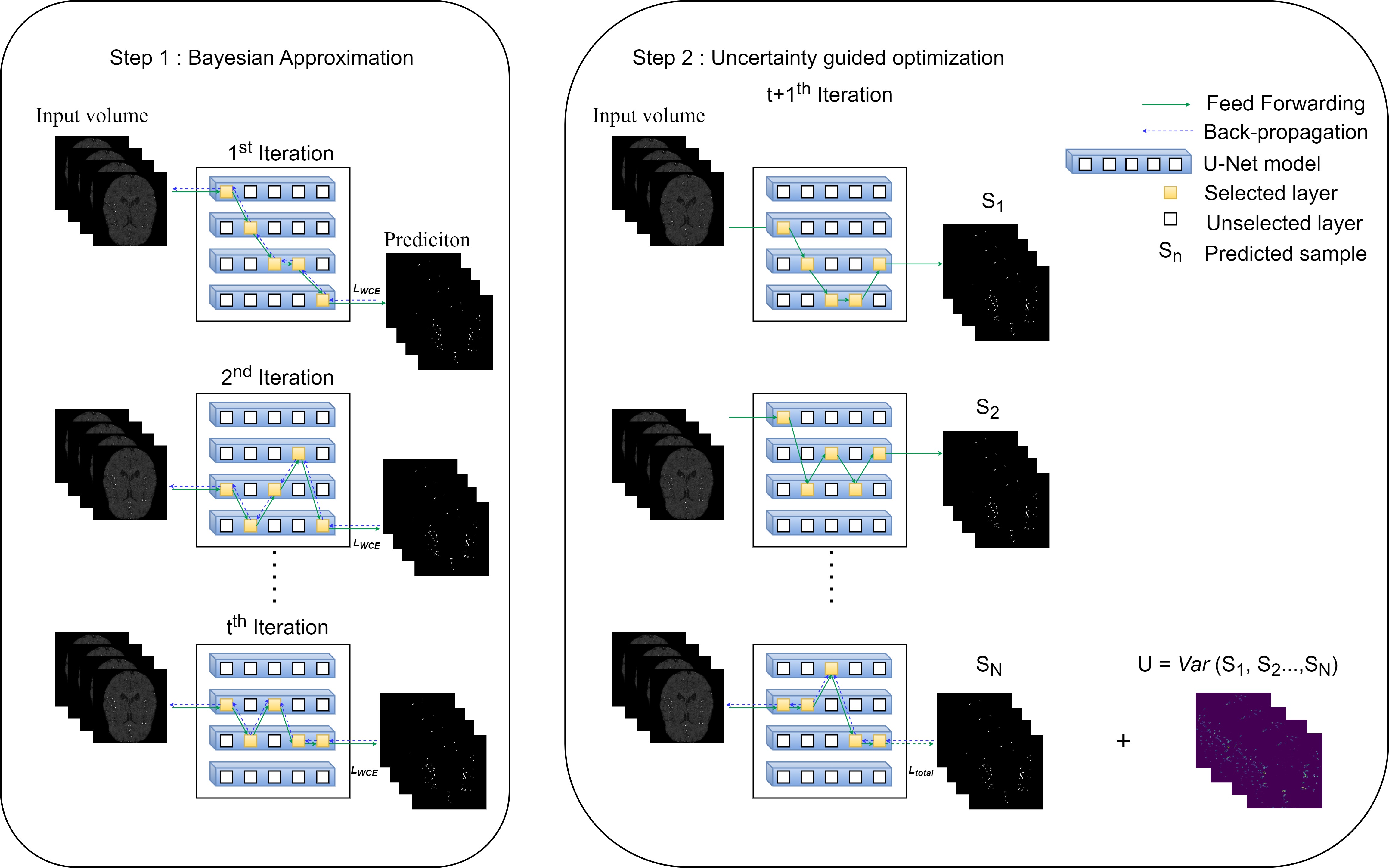}
    \caption{Efficient Ensemble Architecture: Ensemble members are created by selecting the layers randomly at each iteration. In step one the model is trained to minimize cross entropy loss, in step two N samples are generated and their variance is used to estimate $L_U$.}
    \label{fig:Ensemble}
\end{figure}
Thus, if a model architecture consists of $L$ layers, $M^{L}$ different ensemble members can be constructed. If a random variable $P_{l}(m)$ equals $1$, the corresponding layer $l$ of sub-model $m$ is selected. In order to maintain the network architecture, it is important to ensure that a layer is only selected by one sub-model, hence the condition $\sum_{m=1}^{M} P_{l}(m) = 1 $ is fulfilled for all layers $l$. The assignment of a categorical random variable $P_{l}(m)$ to each layer follows a Bernoulli distribution. As claimed in \cite{ref_proc3}, this is equivalent to performing variational inference on the network parameters. As a model with Bernoulli-distributed weights approximates a Gaussian process \cite{ref_gal}, a Bayesian approximation technique can be applied to estimate uncertainty. 
The ensemble models were optimized in two stages: Weighted cross-entropy (WCE) loss is used as loss function in the first step. To account for the high imbalance between foreground (i.e. vessel) and background (i.e. no vessel) voxels, weighted cross entropy loss ($L_{\textit{WCE}}$) was used for optimization with weights of 0.1 for background and 0.9 for foreground. Next, the trained model was used to obtain segmentation maps of training data. From these segmentation maps, voxel-wise uncertainty maps $U_{i}$ for each image $i$ were generated by calculating the variance across $s= 20$ ensemble outputs. In the second step, these uncertainty maps were incorporated into the loss function:  
\begin{equation}
    L_{U}=-\sum_{i}^{N}(U_{i} \otimes y_{i})\log(U_{i} \otimes \hat{y}_{i})
\end{equation}
\begin{equation}
    \label{totalLoss}
    L_{\text{total}} = L_{WCE} + \lambda  L_U
\end{equation}
Where $y_{i}$ and $\hat{y}_{i}$ refer to the ground truth and output segmentation of the network, respectively. By multiplying the obtained uncertainty maps with the ground truth segmentations via the Hadamard product $\otimes$, higher weights were set on uncertain voxels. The balancing parameter {$\lambda$} was set to 10.0 to ensure equilibrium between the two loss components. \\
\textbf{Backbone Model Architectures}
In this work, we developed a 3D U-Net model optimized for cerebrovascular segmentation. Our segmentation model consists of four down-sampling layers consisting of striding convolutions. Each convolutional
block consists of a convolution operation followed by instance normalisation, and Leaky ReLU activation layer. The number of feature maps increases from 32 in the first layer to 64, 128, and 256 in the second, third, and forth layer, respectively. 
We compare our 3D-UNet with two other architectures used in the literature for brain vessel segmentation: (1) DeepVesselNet \cite{ref_deepVN} (DVN), (2) A U-Net with half of the original U-Net parameters but more parameters than ours \cite{ref_redUNET} consisting of double convolution blocks followed by normalization, and  ReLU activation layer. For down-sampling, max-pooling is used. The original architecture uses batch norm but as efficient ensemble failed to converge we used instance norm instead. For all models, we investigate if the use of the ensemble model increases segmentation accuracy with respect to the deterministic backbone model (referred as baseline model in the following).\\
\textbf{Uncertainty Quantification} To obtain uncertainty maps, mean and variance values were calculated across the $s=20$ sample outputs of the ensemble members. To obtain segmentation results, mean values higher than 0.5 were set to 1 while values equal or below 0.5 were set to 0. \\
\textbf{Application to OOD-Data} To explore the ability of the networks to generalize to unseen data, we evaluated baseline and ensemble models on the one in-distribution (ID) and four OOD test datasets. Evaluation metrics and uncertainty maps were compared across these datasets. \\
\textbf{Exploring Uncertainty-based Segmentation Correction} To assess the efficiency of the uncertainty estimations, sparsification plots were used to summarize the degree to which the observed uncertainty aligns with segmentation errors \cite{ref_optflow}, \cite{ref_prior}. 

\section{Experiments}
\subsection{Data and Training Details}
\textbf{Training and Testing Dataset} In this work, the COSTA dataset \cite{ref_COSTA} was used which is available on request for research purposes consisting of MRA images acquired at six hospitals. Two datasets containing 120 images (IXI-Guys and IXI-HH) were used for model training, validation, and testing by splitting these datasets randomly in 86 images for training, 20 for model validation, and 14 for testing. The remaining four datasets (IXI-IOP, ADAM, ICBM, and LocH1) were used for testing on OOD data.  
The OOD-data yielded different voxel-sizes than the training data and were acquired at different scanner types. Moreover, in the ADAM and LocH1 datasets, patients with cerebrovascular diseases were present while all other datasets only consisted of healthy patients. Before network training, bias field correction was applied and images were normalized to values between 0 and 1 using Min-Max normalization. Training was performed on random patches of size 128 × 128 × 80 to optimize memory usage. Non-zero labelled voxel-centered patches ensured a balanced representation of vascular structures. Data augmentation techniques including random flipping, contrast adjustment, and random motion were applied.\\
\textbf{Implementation Details} All networks were implemented in PyTorch (version 2.1.2) and trained using PyTorch Lightning. Number of epochs and batch size  are listed in Table \ref{tab:ensemble_config}. The initial learning rate was set to 0.0001, and Adam optimizer \cite{ref_adam} was used for optimization. The learning rate was reduced by a factor of 0.2 after 30 epochs in case no improvement in training loss was observed. The code repository to facilitate research reproducibility will be available here in the final version. \\
\textbf{Testing Details} Inference was performed on full-sized images. As input to the U-Net must yield an image size divisible by the number of down-sampling operations, padding was applied before inference. The predicted segmentation maps were cropped to restore the original image dimensions. To obtain the final segmentation, the largest connected element was determined and used as the final segmentation. All smaller segmented elements were discarded.\\
\textbf{Hardware} All training processes ran on NVIDIA A100 GPUs, 40GB. 1 GPU was used for baseline and 2 GPUs were used for ensemble model training.
\subsection{Evaluation}
For evaluation, we compare the models based on their accuracy, efficiency in detecting segmentation errors, and computational costs. We compare Precision, Recall, Foreground Dice and clDice percentage across backbone model architectures. 
\section{Results}

\textbf{Evaluation on in-distribution data}  We used 3 sub-models for each efficient ensemble with model configurations summarized in Table \ref{tab:ensemble_config}. Our U-Net architecture was more efficient in training time than the DVN (12 h 20 min vs. 15 h 15 min). Due to memory limitations the batch size for DVN and half-sized U-Net was reduced.

\begin{table}[h!]
\caption{Ensemble models configurations}
    \flushleft
    \resizebox{\textwidth}{!}{

    \begin{tabular}{|c|c|c|c|c|c|c|}
        \hline
        \textbf{Network.} & \textbf{\# Sub-models } & \textbf{\# Layers } & \textbf{\# Parameters} & \textbf{Training Time } & \textbf{Nr.} & \textbf{Batch} \\
        &\textbf{({$M$})}&\textbf{({$L$})}&& \textbf{(hh:mm:ss)}&\textbf{ epochs}&\textbf{size}\\
        \hline
        
        3D U-Net (ours) & 3 & 7 & 5.8M & 12:20:00 & 300 & 16 \\ 
        \hline
        
        DVN & 3 & 15 & 139.8K & 15:15:00 & 300 & 2\\ 
        \hline

        U-Net & 3 & 23 & 67.7M & 25:12:00& 400 & 2\\
        \hline
    \end{tabular}}
\label{tab:ensemble_config}
\end{table}

As displayed in Table \ref{tab:accBaseLine}, baseline and efficient ensemble model of our 3D-UNet architecture was superior to the DVN (clDice and Dice \% of 86.03 and 79.69 for our model vs. 80.21 and 75.85 for DVN) while baseline half-sized U-Net performed best. The efficient ensemble U-Net yielded similar results to ours. For our architecture, the use of the efficient ensemble model with 3 sub-models leads to an increase in clDice and Dice \% to 87.0 and 82.37 while it leads to a decrease in both metrics for the DVN (clDice of 77.55 and Dice of 72.9) and U-Net (clDice of 86.33 and Dice of 83.04)

\begin{table}
\caption{Evaluation metrics for baseline and efficient ensemble model for our network, DVN and U-Net.}
    \flushleft
	\resizebox{0.7\textwidth}{!}{
    \begin{tabular}{|c|c|c|c|c|}
    \hline
        Model type& Sensitivity & Precision  & clDice & Dice  \\
        \hline

        Baseline ours& 93.61  & 69.69 & 86.03 & 79.69
	 \\
        \hline
        Ensemble ours& 92.4 & 74.74 &87.0 & 82.37 \\
        \hline

        Baseline DVN & 94.77  & 63.77 & 80.21 & 75.85
	 \\
        \hline
        Ensemble DVN & \textbf{95.94} & 59.34 &77.55 & 72.9 \\
        \hline
        Baseline U-Net & 93.06  & \textbf{82.89} & \textbf{89.64} & \textbf{87.45}
	 \\
        \hline
        Ensemble U-Net & 94.91 & 74.22 & 86.33 & 83.04 \\
        \hline
        
    \end{tabular}}
    
    \label{tab:accBaseLine}
\end{table} 

\begin{figure}[h!]
    \centering
    \includegraphics[width=0.6\linewidth]{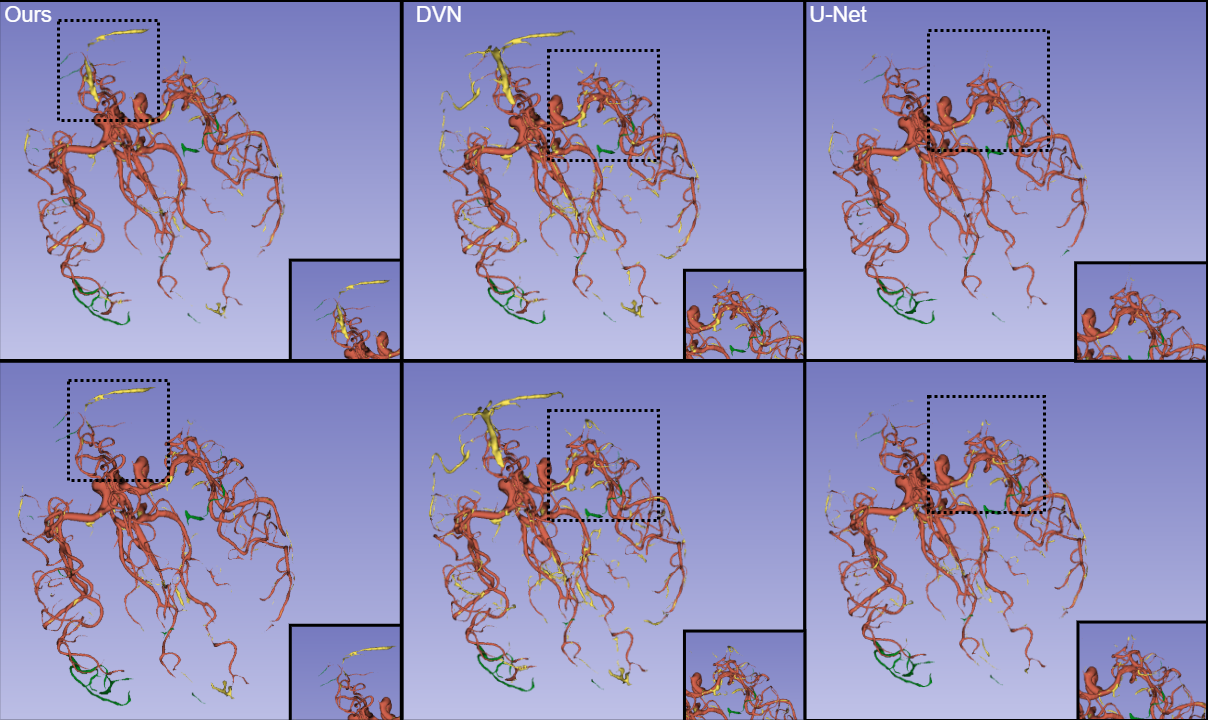}
    \caption{Segmentation error maps of all networks: First row: Baseline results; Second row: Ensemble results. True positive segmented voxels are displayed in red, false positives in yellow, and false negatives in green}
    \label{fig:compare}
\end{figure}

\begin{figure}[h!]
    \centering
    \includegraphics[width=\linewidth]{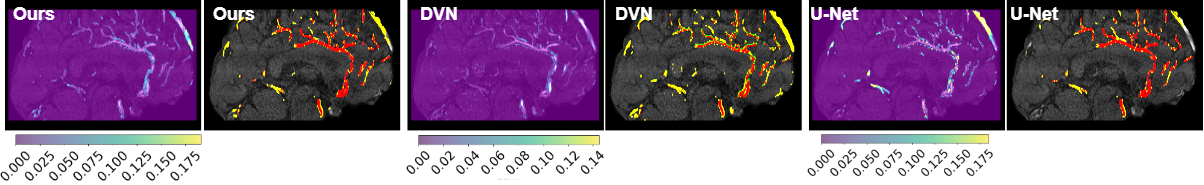}
    \caption{Qualitative results of segmentation and UQ for the three architectures under consideration. 
    Sagittal view of one MRA image. From left to right, we show voxelwise variance (in purple/yellow) and segmentation error overlaid on the input image for our 3D U-Net, DVN and the half-sized U-Net. In the error map, true positives are displayed in red, false positives in yellow, and false negatives in green. }
    \label{fig:variance}
\end{figure}

\textbf{Uncertainty Analysis} In the ID test-set, higher uncertainty was observed near vessel edges and regions adjacent to the skull with intensities similar to those of vessels (Figure \ref{fig:variance}). False Positives and False Negatives were predominantly observed in regions with intricate vessel boundaries and low contrast. These incorrect segmented voxels also yielded high uncertainty indicating the alignment of uncertainty and segmentation errors. These errors might be caused by the differences in vessel thickness across the vasculature leading to imprecise edges. Moreover, regions with vessel-like intensities are rare in the training data which might also cause the uncertainty in these regions. With the help of these uncertainty estimates we can determine which regions would require radiologist's opinion.

\begin{figure}[!h]
\includegraphics[width=\textwidth]{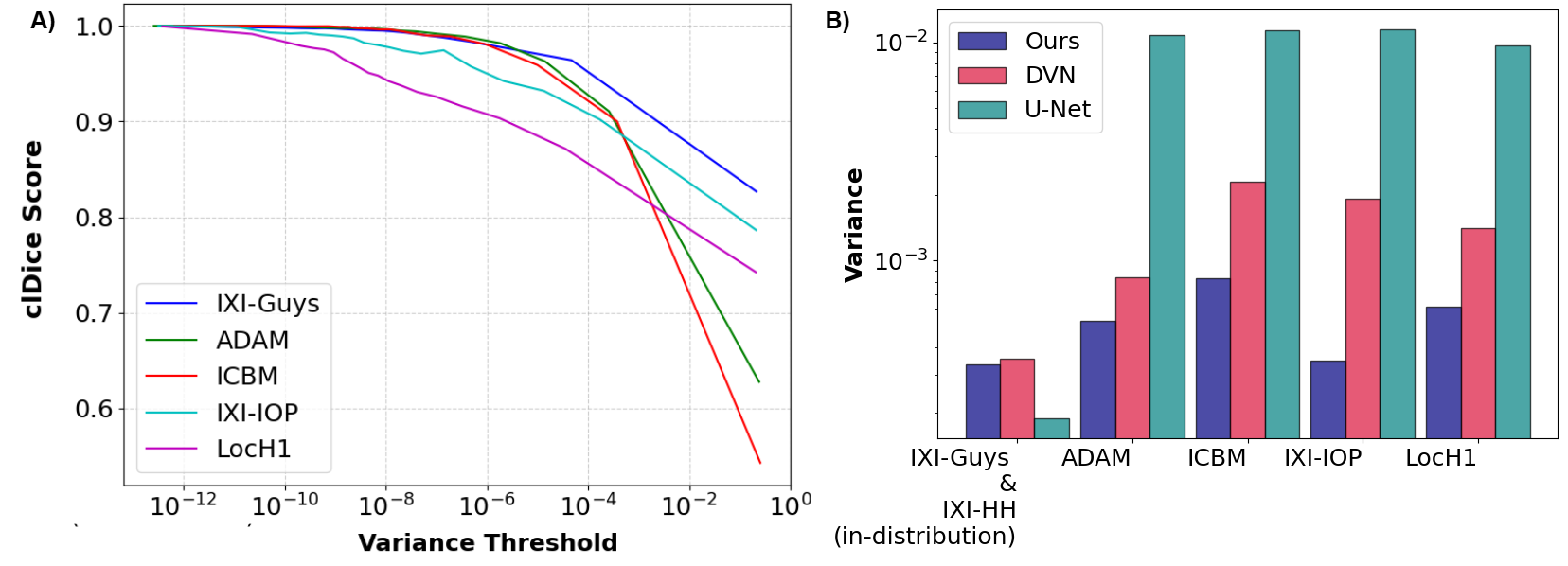} 
\caption{Left (A): clDice vs. variance threshold for our network; Right (B): Average variance of test images for all datasets.}
\label{fig:combined}
\end{figure}

Figure \ref{fig:combined} A (left) shows that the average clDice score of test images increases as we gradually remove the voxels with uncertainty higher than certain thresholds, i.e. when removing voxels with variance larger than $10^{-4}$ would increase the average clDice score from 78 to 91 for the IXI-IOP dataset. Most background voxels yield variance close to zero leading to a clDice of 1. This inverse proportion of accuracy and uncertainty shows that the efficient ensemble model successfully identifies incorrect segmented areas as areas with high uncertainty. Setting variance thresholds whereby foreground voxels exhibiting a certain degree of variance are reassigned to the background can consequently improve segmentation results. However, the best general threshold is image-dependent and would likely need to be determined manually by a radiologist.  
\\
\textbf{Evaluation on OOD-Data}
\begin{table}[h!]
\caption{Evaluation metrics for baseline and efficient ensemble models}
    \flushleft
	\resizebox{0.95\textwidth}{!}{
    \begin{tabular}{|c|c|c|c|c|c|c|c|c|c|c|c|}
    \hline
    &\multicolumn{9}{ c |}{Results Ensemble (Baseline)} \\ \hline
    &\multicolumn{3}{ c |}{ Ours}&\multicolumn{3}{ c |}{DVN}&\multicolumn{3}{ c |}{U-Net}\\
            \hline

        Dataset& Sensitivity & clDice & Dice  & Sensitivity & clDice & Dice  & Sensitivity & clDice & Dice  \\
        \hline

        ADAM& 81.54 (82.47)  & 81.69 (77.45) & 83.1 (81.64) & 93.76 (86.16)  &74.03 (73.92)  &77.73 (79.57) &96.89 (72.11) & 2.7 (77.3) & 1.7(80.55)
	 \\
        \hline
        ICBM& 70.04 (83.43) &71.40 (70.85) & 72.55 (71.02) & 84.93 (90.34)  & 60.5 (48.27) & 58.16 (50.89) & 96.0 (82.95) & 1.47(85.08) & 0.8 (83.04) \\
        \hline

        IXI-IOP & 74.1 (82.03)  & 79.82 (84.16) & 81.86 (82.47) & 90.68 (80.65) & 33.26 (61.16) & 39.07 (64.74) & 96.18 (86.8) & 2.03 (89.78) & 1.35 (88.16)
	 \\
        \hline
        LocH1 & 76.10 (80.98)  & 73.85 (50.93) & 79.76 (74.13) & 87.49 (82.73)  & 54.84(66.23) &65.87 (76.9) &95.74(72.08) &2.31(80.33) &1.4(80.29)  \\
        \hline
        
    \end{tabular}}
    
    \label{tab:accOOD}
\end{table}

\begin{figure}[h!]
    \centering
    \includegraphics[width=0.9\linewidth]{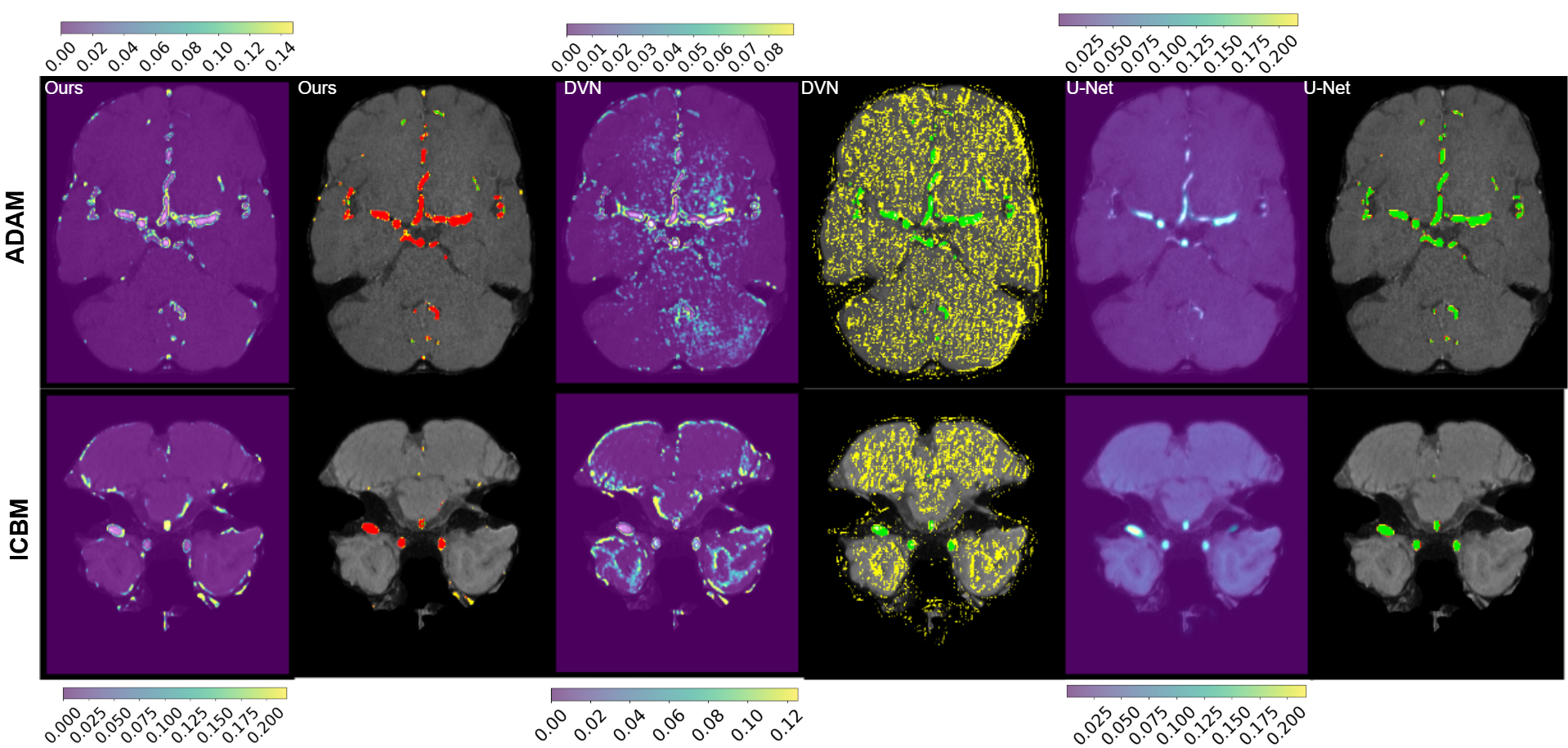}
    \caption{Segmentation results of OOD datasets}
    \label{fig:ood-errormap}
\end{figure}
As displayed in Table \ref{tab:accOOD}, the use of the ensemble model led in all but one OOD-dataset to an increase in accuracy metrics for our model when compared with the baseline model. E.g. for the ADAM dataset, the clDice increased from 77.45 for the baseline model to 81.69 for the ensemble model. In the ICBM dataset, the clDice increased from 47.27 for the DVN baseline to 60.5 for the DVN ensemble model. Interestingly, half-sized U-Net  ensemble did not achieve decent results for the OOD-data also evident by it's higher variance as compared to other two architectures (Figure \ref{fig:combined}B). 
\section{Conclusion}
In this study, we implemented epistemic uncertainty estimation in a cerebrovascular segmentation model using ensemble-based Bayesian approximation.  We demonstrated that our ensemble model was superior not only to the respective deterministic baseline network but also to DeepVesselNet. In the OOD datasets, higher average variance was observed compared with ID data confirming the capability of the uncertainty maps to explain degraded performance. Also for the OOD data out network resulted in superior performance compared with the other included network architectures. Future works should focus on incorporating aleatoric uncertainty and improving the backbone network for better generalization.

\begin{credits}
\subsubsection{\ackname} The authors gratefully acknowledge computing time on the supercomputer JURECA\cite{ref_ack} at Forschungszentrum Jülich under grant no. delia-mp.
\end{credits}


\begin{thebibliography}{8}
\bibitem{ref_article1}
Kay Sin Tan, Jeyaraj Durai Pandian, Liping Liu, Kazunori Toyoda, Thomas Wai Hon Leung, Shinichiro Uchiyama, Sathoshi Kuroda, Nijasri C. Suwanwela, Sanjith Aaron, Hui Meng Chang, Narayanaswamy Venketasubramanian; Stroke in Asia. Cerebrovasc Dis Extra 16 August 2024; 14 (1): 58–75. \url{https://doi.org/10.1159/000538928}

\bibitem{ref_proc1}
F. Taher, A. Mahmoud, A. Shalaby, and A. El-Baz, 
``A review on the cerebrovascular segmentation methods,'' 
in \textit{2018 IEEE International Symposium on Signal Processing and Information Technology (ISSPIT)}, 
2018, pp. 359--364, doi: \url{https://doi.org/10.1109/ISSPIT.2018.8642756}.

\bibitem{ref_article2}
C. Chen, K. Zhou, Z. Wang, Q. Zhang, and R. Xiao, 
``All answers are in the images: A review of deep learning for cerebrovascular segmentation,'' 
\textit{Computerized Medical Imaging and Graphics}, 
vol. 107, p. 102229, 2023. 
Available: \url{https://www.sciencedirect.com/science/article/pii/S0895611123000472}, 
doi: \url{https://doi.org/10.1016/j.compmedimag.2023.102229}.

\bibitem{ref_article3}
X. Chen, Y. Lei, J. Su, H. Yang, W. Ni, J. Yu, Y. Gu, and Y. Mao, 
``A review of artificial intelligence in cerebrovascular disease imaging: Applications and challenges,'' 
\textit{Current Neuropharmacology}, 
vol. 20, no. 7, pp. 1359--1382, 2022, 
doi: \url{10.2174/1570159X19666211108141446}, 
PMID: 34749621, PMCID: PMC9881077.

\bibitem{ref_article4}
A. Singh, S. Sengupta, and V. Lakshminarayanan, 
``Explainable deep learning models in medical image analysis,'' 
\textit{Journal of Imaging}, 
vol. 6, no. 6, p. 52, Jun. 2020, 
doi: \url{10.3390/jimaging6060052}, 
PMID: 34460598, PMCID: PMC8321083.

\bibitem{ref_corr}
A. Singh, S. Sengupta, and V. Lakshminarayanan, 
``Uncertainty and interpretability in convolutional neural networks for semantic segmentation of colorectal polyps,'' 
\textit{Medical Image Analysis}, vol. 60, p. 101619, 2020, 
doi:\url{https://doi.org/10.1016/j.media.2019.101619}.

\bibitem{ref_proc2}
{\"O} . {C}i{\c{c}}ek, A. Abdulkadir, S. S. Lienkamp, T. Brox, and O. Ronneberger, 
``3D U-Net: Learning dense volumetric segmentation from sparse annotation,'' 
in \textit{Medical Image Computing and Computer-Assisted Intervention -- MICCAI 2016}, 
S. Ourselin, L. Joskowicz, M. R. Sabuncu, G. Unal, and W. Wells, Eds. Cham: Springer International Publishing, 2016, pp. 424--432, 
ISBN: 978-3-319-46723-8.

\bibitem{ref_gal}
Y. Gal and Z. Ghahramani, ``Dropout as a Bayesian Approximation: Representing Model Uncertainty in Deep Learning,''
arXiv preprint 	arXiv:1506.02142, 2015. 
Available: \url{https://arxiv.org/abs/1506.02142}.

\bibitem{ref_proc3}
H. J. Lee, S. T. Kim, H. Lee, N. Navab, and Y. M. Ro, 
``Efficient ensemble model generation for uncertainty estimation with Bayesian approximation in segmentation,'' 
arXiv preprint arXiv:2005.10754, 2020. 
Available: \url{https://arxiv.org/abs/2005.10754}.

\bibitem{ref_deepVN}
G. Tetteh, V. Efremov, N. D. Forkert, M. Schneider, J. Kirschke, B. Weber, C. Zimmer, M. Piraud, and B. Menze, 
``Deepvesselnet: Vessel segmentation, centerline prediction, and bifurcation detection in 3-d angiographic volumes,'' 
in \textit{Frontiers in Neuroscience}, vol. 14. pp. 592352, 2020,
doi: \url{10.3389/fnins.2020.592352}

\bibitem{ref_adam}
D. P. Kingma, J. Ba
``Adam: A Method for Stochastic Optimization,''
arXiv preprint arXiv:1412.6980, 2017. 
Available: \url{https://arxiv.org/abs/1412.6980}.


\bibitem{ref_redUNET}
M. Livne, J. Rieger, O. Aydin, T. Utku, A.A. Taha, E.M. Akay, T. Kossen, J. Sobevsky, J. D. Kelleher, K. Hildebrand, D. Frey, and V.I. Madai,  
``A U-Net Deep Learning Framework for High Performance Vessel Segmentation in Patients With Cerebrovascular Disease,'' 
in \textit{Frontiers in Neuroscience}, vol. 13. pp. 592352, 2019,
doi: \url{10.3389/fnins.2019.00097}
\bibitem{ref_COSTA}
``L. Mou, J. Lin, Y. Zhao, Y. Liu, S. Ma, and J. Zhang, 
COSTA: A Multi-Center TOF-MRA Dataset and a Style Self-Consistency Network for Cerebrovascular Segmentation,''
in \textit{IEEE Transactions on Medical Imaging}, vol. 43. pp. 4442 - 4456, 2024,
doi: \url{10.1109/TMI.2024.3424976}
\bibitem{ref_optflow}
E.~Ilg, \"{O}.~\c{C}i\c{c}ek, S.~Galesso, A.~Klein, O.~Makansi, F.~Hutter, and T.~Brox,  
``Uncertainty estimates and multi-hypotheses networks for optical flow,''  
preprint arXiv:1802.07095, 2018. [Online]. Available: \url{https://arxiv.org/abs/1802.07095}
\bibitem{ref_prior}  
Kamil Ciosek, Vincent Fortuin, Ryota Tomioka, Katja Hofmann, and Richard E. Turner.  
\textit{Conservative Uncertainty Estimation By Fitting Prior Networks}.  
International Conference on Learning Representations, 2020.  
\url{https://api.semanticscholar.org/CorpusID:209319793}
\bibitem{ref_ack}
Jülich Supercomputing Centre. (2021). JURECA: Data Centric and Booster Modules implementing the Modular Supercomputing Architecture at Jülich Supercomputing Centre Journal of large-scale research facilities, 7, A182. \url{http://dx.doi.org/10.17815/jlsrf-7-182}

\end{thebibliography}
\end{document}